\documentclass[conference,a4paper]{IEEEtran}
\IEEEoverridecommandlockouts

\usepackage[hidelinks]{hyperref}
\usepackage[cmex10]{amsmath}
\usepackage{amssymb,amsfonts}
\usepackage{dblfloatfix}

\usepackage[ruled,vlined]{algorithm2e}
\usepackage{graphicx}
\graphicspath{{Figures/PDF/}{Figures/PNG/}}

\usepackage{booktabs}
\usepackage{siunitx}
\usepackage[numbers,compress]{natbib}
\usepackage{texnames}
\usepackage{bm,bbm}
\usepackage{orcidlink}
\usepackage{array}
\usepackage{placeins}
\usepackage{float}
\usepackage{dblfloatfix}

\author{%
\IEEEauthorblockN{Abira Sengupta\orcidlink{0000-0002-6867-3362}}
\IEEEauthorblockA{\textit{University of Otago}\\
Dunedin, New Zealand\\
sengupta.abira0609@gmail.com}
\and
\IEEEauthorblockN{Ayoti Banerjee\orcidlink{0009-0007-0374-1559}}
\IEEEauthorblockA{\textit{Indian Statistical Institute}\\
700108, West Bengal, India\\
ayoti.banerjee@gmail.com}
\and
\IEEEauthorblockN{Sarbani Palit\orcidlink{0000-0002-4105-6452}}
\IEEEauthorblockA{\textit{Indian Statistical Institute}\\
700108, West Bengal, India\\
sarbanip@isical.ac.in}
\and
\IEEEauthorblockN{Brendon Woodford\orcidlink{0000-0001-5202-9700}}
\IEEEauthorblockA{\textit{University of Otago}\\
Dunedin, New Zealand\\
brendon.woodford@otago.ac.nz}
}

\makeatother

\begin{document}

 \title{Blackening Cryosphere: Revealing Hotspot Shifts and HGB-Based Forecasting of Absorbing Aerosol Threats over the Himalayan Frozen Frontiers}


\maketitle

\begin{abstract}

Black carbon and mineral dust are key absorbing aerosols that influence atmospheric radiation and increasingly threaten global cryospheric stability. This study examines the long-range transport and seasonal variability of these aerosols over Pakistan and their movement toward the western Himalayas. Using satellite-derived Absorption
Aerosol Optical Depth (AAOD) data from 2019 to mid-2025, we analyse their spatiotemporal behaviour across Pakistan's urban lowlands and high-altitude regions. Fifteen-day aggregated AAOD fields are used to track seasonal transport into glaciated terrain, where deposited aerosols can darken snow and ice and accelerate melt. For high-AAOD events, a probabilistic forecasting approach based on machine learning (ML) was created. Using geographical, seasonal and lagged indicators, a Histogram-based Gradient Boosting classifier was trained to predict AAOD exceedance one step in advance. ROC-AUC, PR-AUC, and the Brier score were used to assess the model's performance. The results showed high predictive capacity and good probability calibration, with values of 0.791, 0.269 and 0.028, respectively. Forecasts show that areas adjacent to the Himalayan glaciers, consistently exhibit the highest probability of increasing AAOD, signalling an elevated risk of aerosol-induced snowmelt.

\end{abstract}

\begin{IEEEkeywords}
	Absorption Aerosol Optical Depth, Black Carbon, Histogram-based Gradient Boosting, Glacier melting.
\end{IEEEkeywords}

\section{Introduction}

Satellite-derived aerosol optical metrics are widely used to infer ground-level particulate matter and complement surface monitoring. Strong empirical relationships between columnar Aerosol Optical Depth (AOD) and surface $\mathrm{PM}_{2.5}$ under varying meteorological conditions have underpinned models linking satellite observations to near-surface pollution \cite{kumar2007empirical}. Beyond total AOD, Absorption Aerosol Optical Depth (AAOD) has emerged as a more sensitive proxy for combustion-related aerosols, effectively capturing variability in light-absorbing particles such as black carbon and mineral dust from fossil fuel and biomass burning \cite{li2021satellite}. High-altitude observations in the central Himalayas show that although black carbon constitutes only about $\sim5\%$ of aerosol mass, it dominates radiative absorption, illustrating why AAOD isolates absorbing signals that are often masked in extinction-based AOD by scattering particles \cite{pant2006aerosol}. At broader scales, OMI satellite retrievals over China reveal clear winter–spring peaks in AAOD \cite{mao2024spatiotemporal}, reinforcing its value as a regional and seasonal indicator of combustion-related PM\textsubscript{2.5}, especially in regions with limited ground-based monitoring or incomplete emissions inventories \cite{li2021satellite}.
 
Meteorological conditions strongly regulate $\mathrm{PM}_{2.5}$ through their control on dispersion, dilution, aggregation and retention \cite{zhao2015influence}. Wind governs its transport across local to trans-boundary scales \cite{qu2015effect}, while high relative humidity promotes secondary aerosol formation and hygroscopic growth, elevating concentrations \cite{chen2016simulations}. By scattering and absorbing solar radiation, $\mathrm{PM}_{2.5}$ alters the atmospheric radiative balance and temperature structure \cite{yang2020impacts}. Accumulation in the lower troposphere can intensify near-surface heating and prolong pollution episodes \cite{dey2008probable}, with even small summer-time variations affecting surface temperatures \cite{im2012summertime}. Together, these processes highlight aerosols as both air pollutants and significant climate forcers.

Absorbing aerosols increase solar absorption, speed up snowmelt, by intensifying atmospheric heating and lowering the albedo of snow and ice surfaces and increase cryospheric risks such glacial lake outburst floods \cite{mustafi2023prediction, nair2013black}. Building on this, the study examines intensified aerosol absorption over high-altitude areas and tracks the spatiotemporal transport of absorbing aerosol hotspots from Pakistan’s urban–industrial lowlands to the western Himalayan cryosphere. We monitor AAOD migration and accumulation across the Himalayas using six years of 15-day averaged satellite AAOD. We then use a histogram-based gradient boosting (HGB) classifier to probabilistically forecast time-dependent high-absorption occurrences (AAOD $\geq$ 0.5). The forecasting methodology uses lag AAOD values from prior timesteps to explicitly reflect temporal persistence and enable real one-step-ahead prediction of future high-absorption events, in contrast to methods based only on geographical and seasonal variables. Given previous evidence of aerosol-induced atmospheric warming and snow-albedo feedbacks, we further investigate whether enhanced aerosol absorption is linked to surface air temperature fluctuation. This integrated approach identifies short-term aerosol hotspots over the Himalayan cryosphere and assesses their potential role in surface warming and accelerated snowmelt.

\section{Study Area}
\begin{figure}[!th]
      \centering
        \includegraphics[width=1\columnwidth]{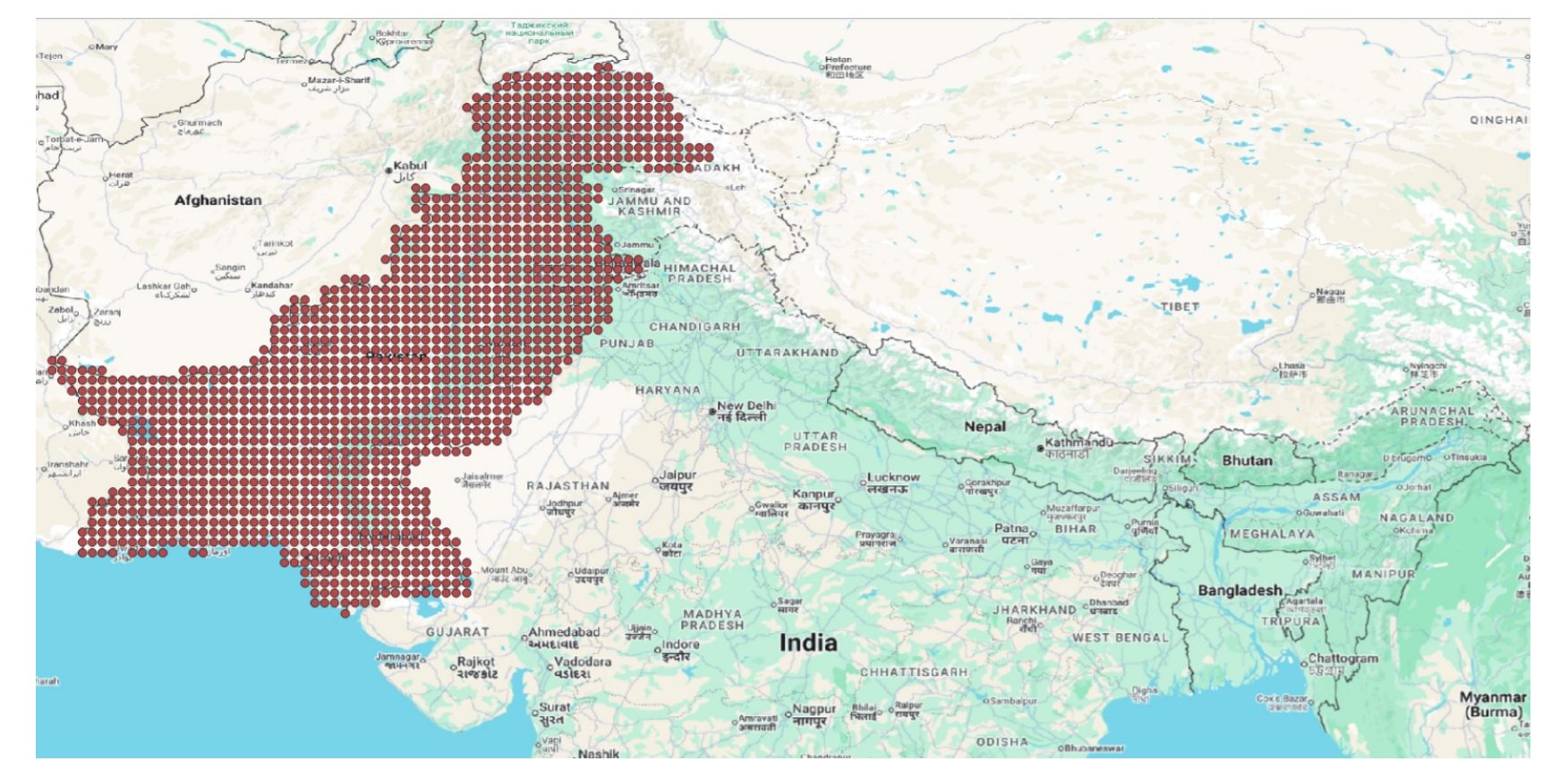} 
     \caption{Geographic extent of the study area encompassing Pakistan and the northern high-altitude regions adjoining the western Himalayas, Karakoram and Hindu Kush ranges. The spatial data points are marked in red.}
       \label{fig:study_area}
   \end{figure}

This study examines Pakistan, a region heavily impacted by anthropogenic $\mathrm{PM}_{2.5}$ emissions from densely populated urban centers. Major cities such as Lahore, Karachi, Faisalabad and Islamabad experience high levels of industrial activity, vehicular emissions and biomass burning, contributing significantly to aerosol loading in the atmosphere. Figure~\ref{fig:study_area} illustrates the geographical extent of the study area, including Pakistan’s major emission hotspots and its northern high-altitude regions that border the western Himalayas, Karakoram and Hindu Kush ranges.

Elevated absorbing aerosol levels over Pakistan have been documented \cite{syed2024tale} and are advected northward into the Himalayan cryosphere by prevailing winds. Long-range transport from the Indo-Gangetic Plains and northwestern South Asia is well established \cite{chen2023carbonaceous}, with black carbon identified as a major driver of cryospheric change. To evaluate these impacts, we use AAOD at 342 nm as a proxy for fine absorbing particles. As the concentration of fine absorbing particles increases, AAOD at 342 nm generally rises \cite{shin2019absorption}.

\section{Data and Methods}

This study uses AAOD at 342 nm from the OMAERUVd v003 Level-3 product (OMI, Aura satellite, NASA), providing global daily aerosol data and its absorption
properties, at \SI{1}{\degree} spatial resolution. Complementary data include Surface Air Temperature from the FLDAS2 Central Asia model, at \SI{0.01}{\degree} spatial resolution (daily values in Kelvin). Both datasets were obtained through the NASA Giovanni portal. Data retrieval was performed for the temporal range spanning from January 1, 2019, to June 30, 2025. To reduce noise and seasonal variability while preserving short-term trends, data were aggregated into 15-day averages. Subsequently, the data were converted to point vectors in QGIS \cite{QGIS_software} to enable spatial–temporal sampling of variables. These point layers were extracted as CSV files for statistical and machine learning analyses.

\begin{figure}[!th]
      \centering
       \includegraphics[width=0.55\textwidth]{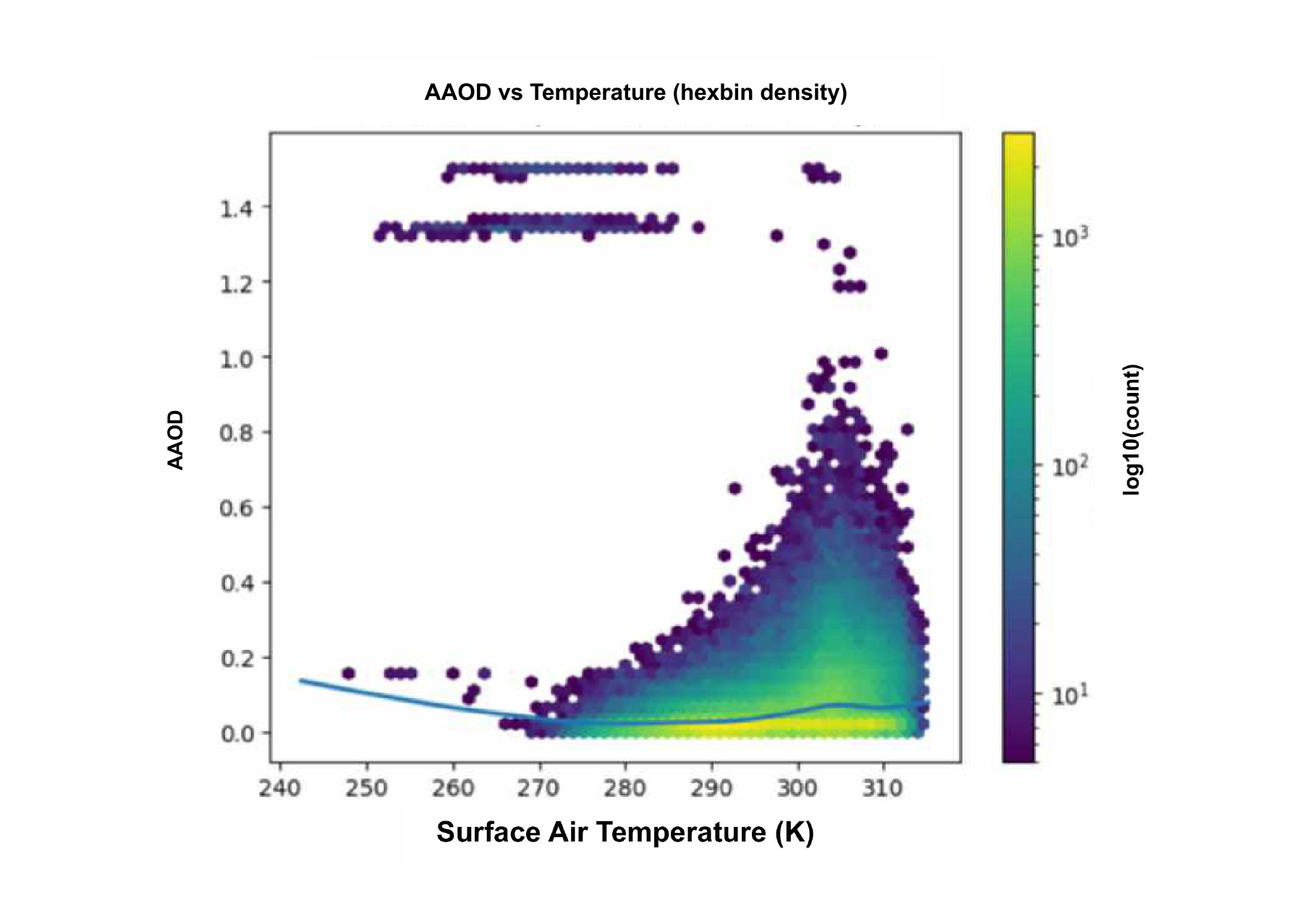} 
      \caption{Hexbin density plot of AAOD versus surface air temperature (in Kelvin). Each hexagonal bin represents the number of observations on a logarithmic scale (color bar).}
       \label{fig:Figure}
   \end{figure}

Then it was examined whether elevated AAOD concentrations are associated with increases in surface air temperature using a scatter matrix (Figure~\ref{fig:Pairwiserelationships}) and a hexbin density plot (Figure~\ref{fig:Figure}). In addition, spatial variance maps of AAOD were generated at 15-day intervals from 1 January 2019 to 30 June 2025 to monitor the temporal migration of aerosol hotspots and depict the evolving patterns of AAOD variability. Six representative maps highlighting key spatial changes are shown in Figure~\ref{fig:PMspreading}.

HGB classifier was employed to model and predict high aerosol absorption events. The model builds an ensemble of decision trees sequentially, but unlike standard gradient boosting, it first discretises continuous predictors into bins, substantially improving computational efficiency without loss of accuracy. Implemented as a binary probabilistic classifier with AAOD $\geq$ 0.5 defining high events, it produces calibrated probabilities for direct spatial risk forecasting.

In addition to spatial coordinates and seasonal temporal features, lagged AAOD values were incorporated as predictors. Specifically, AAOD at previous timesteps ($t\!-\!1$, $t\!-\!2$, and $t\!-\!3$) was used to capture temporal persistence in aerosol absorption and to enable true one-step-ahead forecasting. Model training utilised the first 156 timesteps, while subsequent timesteps were reserved for testing to ensure realistic predictive evaluation; missing values were addressed using median imputation. After evaluation, the model was refitted using all available data up to $t=156$ to generate a one-step-ahead forecast for $t=157$. The final configuration employed a maximum tree depth of 6, a learning rate of 0.08 and 400 boosting iterations.

Model performance was evaluated using three complementary probabilistic metrics: Receiver Operating Characteristic Area Under the Curve (ROC-AUC), Precision–Recall Area Under the Curve (PR-AUC) and Brier Score. ROC-AUC was used to examine the model's overall discriminative power, PR-AUC to evaluate performance under class imbalance by highlighting detection of high-AAOD events, and Brier Score to test accuracy of the projected probabilities.

\section{Results and Discussion}

\begin{figure}[!th]
      \centering
       \includegraphics[width=0.8\columnwidth]{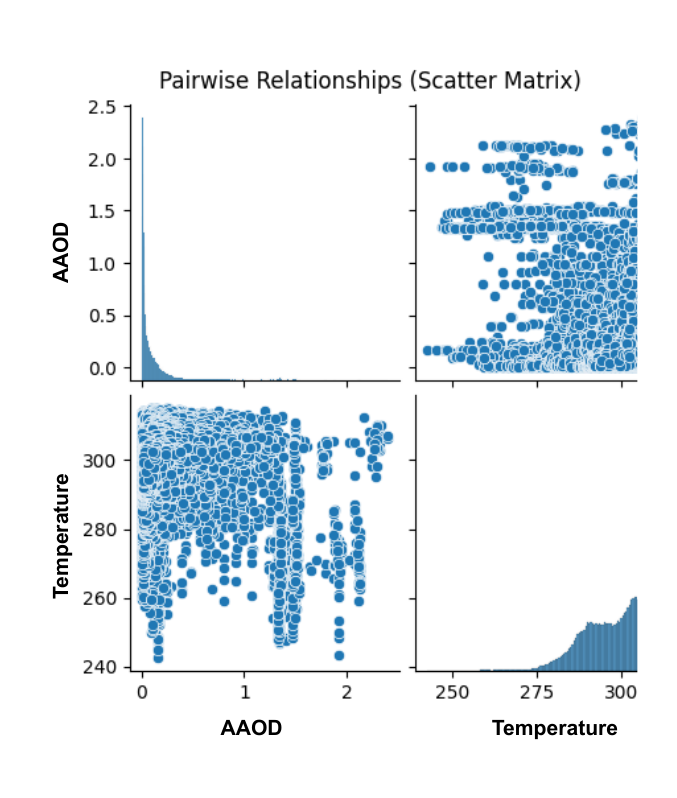}
      \caption{AAOD and surface air temperature (in Kelvin) scatter matrix. }
       \label{fig:Pairwiserelationships}
   \end{figure}

The scatter matrix reveals only a weak relationship between AAOD and surface air temperature, with no clear linear trend (Figure~\ref{fig:Pairwiserelationships}). This is supported by correlation analysis, which shows mild monotonic associations (Spearman $\approx 0.33$, Kendall Tau $\approx 0.23$). These results indicate that while AAOD at 342 nm exerts some influence on temperature, it is not the sole controlling factor for temperature rise. The AAOD distribution is right-skewed, with most values between 0 and 0.5, whereas temperature exhibits a multimodal structure with peaks near 255, 290 and 305 K. 
Hexbin analysis further shows that most observations cluster between 270–310 K and AAOD$<$0.5, although elevated AAOD levels ($>$ 0.5) are preferentially associated with temperatures above 290 K. Horizontal banding in the scatter plots suggests discretisation effects related to measurement resolution or rounding.
 
\begin{figure*}[!h]
    \centering
    \includegraphics[width=0.95\textwidth]{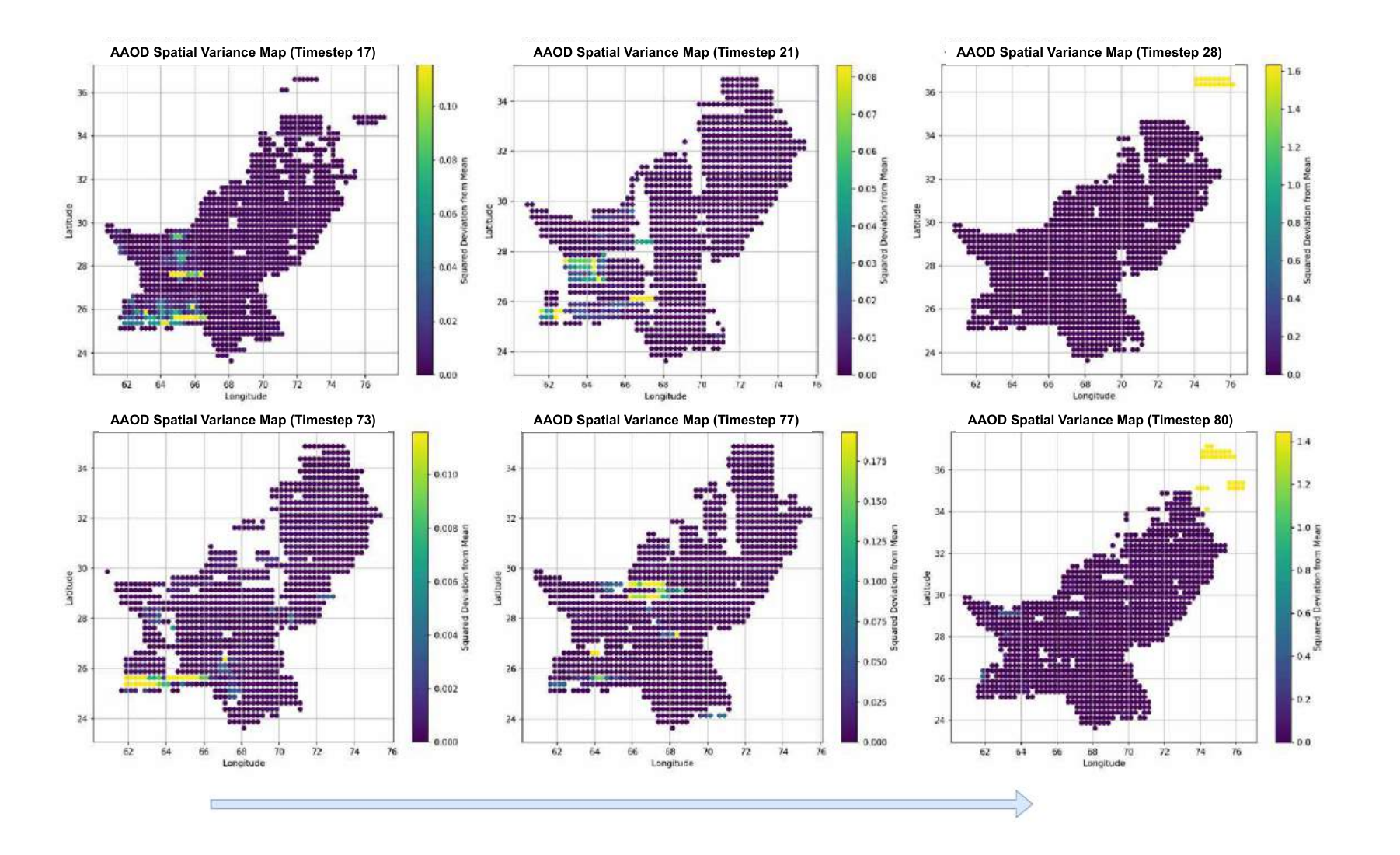}
    \caption{Spatial variance maps of AAOD across Pakistan: Timestep 17 (01–15 September 2019), Timestep 21 (01–15 November 2019), Timestep 28 (16–29 February 2020), Timestep 73 (01–15 January 2022), Timestep 77 (01–15 March 2022), and Timestep 80 (16–30 April 2022).}
    \label{fig:PMspreading}
\end{figure*}

Seasonal dynamics and transport pathways are clarified using AAOD spatial variance rasters. At Timestep 17 (01–15 September 2019), spatial variance maps (Figure~\ref{fig:PMspreading}) show that southwest and western Pakistan exhibit the highest AAOD variance, indicating strong local emissions and large pollution fluctuations likely driven by industrial activity, transport and boundary-layer dynamics. While the southwest continues to show significant variability, by Timestep 21 (01–15 November 2019), the trend of high variance starts to move eastward and somewhat northward. The start of cooler weather and shifting wind patterns in late fall may have an impact on this transitional movement. At Timestep 28 (16–29 February 2020), representing the late-winter period, AAOD values are markedly elevated over the Himalayan region, while the adjacent lowland areas exhibit comparatively subdued levels.

In early 2022, new hotspots appear at Timestep 73 (1–15 January) in the southwest. By Timestep 77 (1–15 March), AAOD variability shifts northward, with hotspots emerging in central and northeastern regions, likely reflecting the transition from winter to pre-monsoon circulation that favours long-range transport from the plains toward the Himalayas. By Timestep 80 (16–30 April), variance accumulates strongly over northern and northeastern areas near the mountains, indicating that aerosols from densely populated lowlands are not only transported but also concentrated in high-altitude, cryospherically sensitive zones, highlighting the need for coordinated regional emission control and monitoring.

\begin{figure}[htbp]
    \centering
    \includegraphics[width=1\linewidth]{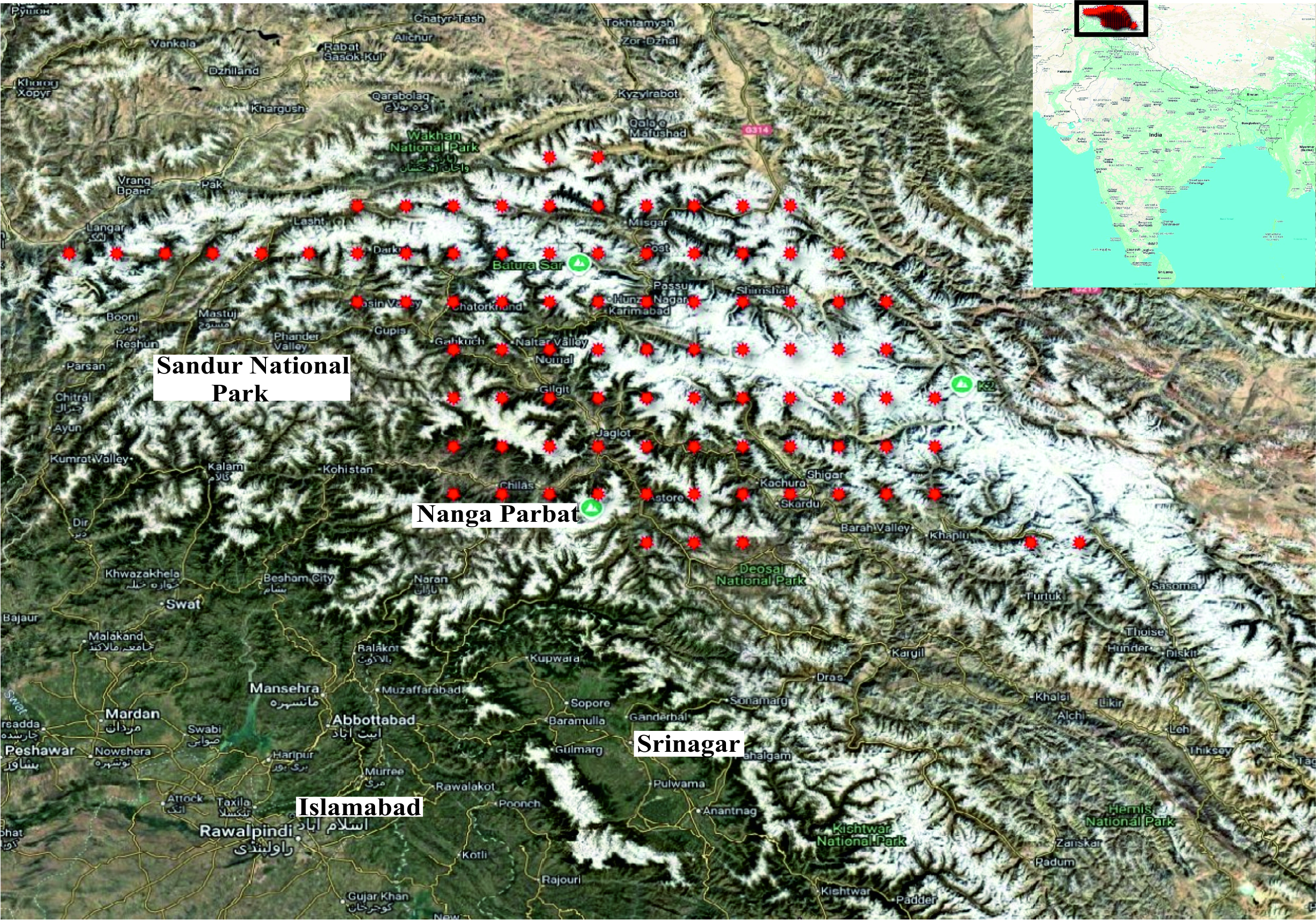}
    \caption{One-step-ahead probabilistic forecast of AAOD exceedance (AAOD $\ge 0.5$) for timestep 157 (1–15 July 2025). Red stars indicate locations with very high risk of exceedance, predominantly over glaciated regions of the western Himalayas.}
    \label{fig:riskmap}
\end{figure}

The HGB model produced a one-step-ahead probabilistic forecast for $t=157$, generating spatially explicit probabilities of AAOD exceedance (AAOD $\ge 0.5$) across the study area. Unlike the earlier seasonal-only formulation, the model incorporates temporal persistence through lagged predictors (AAOD at $t-1$, $t-2$ and $t-3$), along with spatial coordinates and seasonal Fourier terms. Predicted probabilities were classified as low ($p<0.30$), moderate ($0.30 \le p < 0.60$), high ($0.60 \le p < 0.85$), and very high ($p \ge 0.85$). Results for $t=157$ show predominantly low risk over most of Pakistan, with high and very high risk (depicted by red stars in Figure~\ref{fig:riskmap}) concentrated in the far north-northeast glaciated region.

Model performance was assessed using a time-aware split (training on earlier timesteps and testing on later ones). The HGB model achieved a ROC-AUC of 0.8429, a PR-AUC of 0.2595, and a Brier score of 0.0286. The ROC-AUC indicates strong discrimination between exceedance and non-exceedance events, while the PR-AUC confirms reliable detection of rare high-AAOD cases under class imbalance. The low Brier score reflects robust probabilistic accuracy, supporting the use of predicted probabilities for spatial risk assessment.

The pronounced hazard in the northern and northeastern sectors aligns with pre-monsoon and summer circulation regimes that intensify long-range aerosol transport into the Himalayan region \cite{saikawa2019air}. Elevated AAOD over high-altitude terrain promotes surface darkening of snow and ice through deposition of light-absorbing aerosols such as mineral dust and black carbon, thereby enhancing radiative forcing and accelerating melt processes \cite{nair2013black}. Consistent with this mechanism, glacier lake outburst floods (GLOFs) have been repeatedly documented in valleys of this region during April–July, underscoring the climatic and cryospheric vulnerability of these landscapes \cite{mustafi2025assessment}.









\section{Conclusions and Future Outlook}
Using multi-year satellite observations, this study examined the spatiotemporal behaviour of AAOD over Pakistan and adjacent high-altitude regions of the western Himalayas. The AAOD distribution is strongly right-skewed, with most values below 0.5, yet exceedances above 0.5 are closely linked to surface air temperatures exceeding 290 K. Complementing this finding, spatial variance analysis reveals a recurrent pre-monsoon northward shift in aerosol activity from the arid, urbanised southwest, driven by regional circulation that favours long-range transport. During this phase, elevated AAOD over glacial terrain suggests enhanced aerosol deposition, likely reducing surface albedo, amplifying radiative absorption, and accelerating snow and ice melt, thereby increasing cryospheric vulnerability \cite{mustafi2025assessment}.

 A probabilistic machine-learning framework was developed to forecast high aerosol absorption events. A histogram-based gradient boosting (HGB) classifier was trained using spatial coordinates, seasonal terms, and lagged AAOD values, enabling real one-step-ahead predictions that combine temporal persistence with spatial structure. The model achieved a ROC-AUC of 0.8429, a PR-AUC of 0.2595, and a Brier score of 0.0286, indicating strong discrimination between exceedance and non-exceedance events, meaningful skill in identifying rare high-AAOD cases, and robust probabilistic reliability. Projections show that High and Very High risk zones are primarily concentrated in the northern and northeastern Himalayan foothills (Figure~\ref{fig:riskmap}).

While this study advances understanding of absorbing aerosols, it is limited by uncertainties associated with satellite-derived AAOD, particularly over complex terrain and its indirect linkage to surface impacts. Stronger assessments would benefit from combining satellite observations with ground-based measurements and atmospheric modelling. Reliance on AAOD at 342 nm also restricts aerosol characterisation; using multi-wavelength data would improve source attribution and predictive reliability. While 15-day averaging provides a useful and stable representation of regional AAOD patterns, the use of higher temporal resolution data would further improve the ability to capture rapid and short-term aerosol variability. Further work is needed to more clearly quantify how AAOD affects surface temperature and glacier melt in the Himalayas.



\small
\bibliographystyle{IEEEtranN}
\bibliography{references}

\end{document}